\documentclass[lettersize,journal]{IEEEtran}
\usepackage{amsmath,amsfonts}
\usepackage{array}
\usepackage[caption=false,font=normalsize,labelfont=sf,textfont=sf]{subfig}
\usepackage{textcomp}
\usepackage{stfloats}
\usepackage{url}
\usepackage{graphicx}
\usepackage{cite}
\graphicspath{{figures/}}
\usepackage{booktabs}
\usepackage{multirow}
\usepackage{xcolor}
\usepackage{balance}
\usepackage[colorlinks,linkcolor=blue,citecolor=blue,urlcolor=blue]{hyperref}

\newcommand{\eg}{\textit{e.g.}}

\newcommand{\dB}{\,\text{dB}}
\newcommand{\ms}{\,\text{ms}}

\begin{document}

\title{ANVIL: Accelerator-Native Video Interpolation\\ via Codec Motion Vector Priors}

\author{Shibo~Liu%
\thanks{S. Liu is with the College of Science, North China University of
Science and Technology, Tangshan 063210, China
(e-mail: spencerliu@stu.ncst.edu.cn).}%
}


\maketitle
\makeatletter
\def\ps@IEEEtitlepagestyle{%
  \def\@oddfoot{\footnotesize This work has been submitted to the IEEE
    for possible publication. Copyright may be transferred without notice,
    after which this version may no longer be accessible.\hfill}%
  \def\@evenfoot{}}
\makeatother
\thispagestyle{IEEEtitlepagestyle}

\begin{abstract}
Real-time 30-to-60\,fps video frame interpolation on mobile neural processing units (NPUs) requires each synthesized frame within 33.3\,ms. We show that mainstream flow-based video frame interpolation faces three structural deployment barriers on mobile NPUs: spatial sampling operators exceed the frame budget or lack hardware support, iterative flow refinement collapses under 8-bit integer post-training quantization, and memory-bound operators dominate the inference graph. ANVIL addresses these barriers by reusing motion vectors from the H.264/AVC decoder to prealign input frames, removing learned optical flow, spatial sampling, and iterative accumulation from the accelerator graph. The remaining residual is refined by a convolution-dominated network composed almost entirely of compute-bound operators. On a Snapdragon 8 Gen\,3 device, ANVIL achieves 12.8\,ms 1080p inference at 8-bit integer precision; an open-source Android player sustains 28.4\,ms median end-to-end latency over 30-minute continuous playback. Per-operator causal analysis identifies quantized accumulation on recurrent flow states as a key mechanism behind integer quantization failure in iterative methods. The current design targets H.264/AVC playback with decoder-exposed motion vectors.
\end{abstract}


\section{Introduction}
\label{sec:intro}

\IEEEPARstart{V}{ideo} frame interpolation (VFI) synthesizes intermediate frames for frame rate upconversion, slow-motion generation, and video enhancement. Recent methods---RIFE~\cite{rife2022} with iterative intermediate flow estimation, IFRNet~\cite{ifrnet2022} with joint feature--flow refinement, AMT~\cite{amt2023} with all-pairs multi-field transforms---have advanced interpolation quality substantially. Several works have explored lightweight or compressed variants for resource-constrained settings~\cite{cdfi2021}, and mobile video deployment has been studied for related tasks such as neural video compression~\cite{mobilenvc2024}. However, systematic deployment evidence for mainstream VFI pipelines at full resolution on public mobile NPU stacks remains scarce, particularly under operator-compatibility and practical W8A8 constraints. Offloading interpolation to client devices can reduce server compute for streaming platforms and enables privacy-sensitive offline use cases such as mobile gallery slow-motion; in both scenarios, INT8 quantization robustness and operator compatibility on mobile NPUs become first-class design constraints.

To understand the deployment gap, we systematically benchmarked operators drawn from 9 representative VFI methods~\cite{rife2022,ifrnet2022,amt2023,emavfi2023,vfiformer2022,uprnet2023,m2m2022,abme2021,flavr2023} on two mobile NPU platforms from different vendors. The results reveal that deployment bottlenecks stem not from model size or FLOPs, but from structural operator--accelerator mismatch, manifesting in three ways:

\textbf{(1)~\texttt{grid\_sample} latency is prohibitive.}
The bilinear sampling operation central to flow-based warping~\cite{spatialtransformer2015} requires irregular memory access patterns that conflict with NPU architectures optimized for regular tensor computation. At 1080p, a single \texttt{grid\_sample} exceeds the entire 33.3\,ms frame budget on one platform and is unsupported on the other, affecting 7 of the 9 evaluated methods.

\textbf{(2)~Iterative flow refinement collapses under INT8 quantization.}
Mobile NPUs are optimized for integer arithmetic~\cite{xu2025npu}, with FP16 throughput 3--5$\times$ lower in our measurements (Table~\ref{tab:latency}); for the target 1080p 30$\to$60 setting, FP16 exceeds the latency deadline for every tested model, making W8A8 the practical operating point. However, methods that accumulate flow estimates across iterative stages~\cite{rife2022,ifrnet2022} suffer severe quality degradation under W8A8 post-training quantization. Per-operator instrumented analysis traces this to the \texttt{Add} operator on recurrent flow states, which amplifies quantization error across iterations.

\textbf{(3)~Memory-bound operations dominate the inference graph.}
Per-operator profiling reveals that convolutions---the only compute-bound, INT8-accelerable operations---account for merely 5\% of inference cycles in a representative flow-based architecture (RIFE~\cite{rife2022}), with the remaining 95\% consumed by memory-bound operations that benefit minimally from INT8 acceleration.

These barriers recur across flow-, kernel-, and attention-based VFI pipelines in our benchmark, indicating a structural problem rather than a model-specific limitation. The findings point to a unified conclusion: the deployment bottleneck lies in the \emph{operator vocabulary}, not in model capacity. Restricting the inference graph to the operator subset that NPUs execute efficiently---standard convolutions and simple pointwise operations, while avoiding \texttt{grid\_sample}, attention-heavy blocks, and recurrent flow accumulation---would resolve all three barriers simultaneously. The key observation enabling this restriction is that the H.264 decoder already computes per-block motion vectors (MVs) during decoding~\cite{h264standard}; these MVs serve as a zero-learnable-parameter motion prior. Prealigning input frames with spatially smoothed MVs reduces the interpolation problem to small-residual prediction, whose limited output dynamic range permits an NPU-friendly residual network well matched to INT8 execution characteristics. This design therefore targets H.264 playback scenarios where MV side-data is available.

Building on this principle, our deployment design decomposes mobile VFI into codec-side prealignment and NPU-side residual refinement. Codec MVs are spatially smoothed and used to prealign input frames (CPU + GPU); an NPU-friendly residual network dominated by standard convolutions and simple pointwise operations, with batch normalization folded into convolution weights at deploy time, handles the remaining prediction on the NPU.

Our contributions span deployment diagnosis, operator-constrained co-design, and system-level validation:

\begin{enumerate}
  \item \emph{Deployment diagnosis.} Through per-operator instrumented quantization, we identify the Add-on-recurrent-state pattern as a key mechanism behind INT8 quality collapse in iterative flow VFI, verified on two methods (RIFE, IFRNet) via ORT and additionally confirmed on QNN for RIFE. A systematic operator-level benchmark quantifies latency and compatibility barriers for 9 VFI methods on two NPU platforms (Sec.~\ref{sec:int8}).
  \item \emph{Operator-constrained co-design.} We show that decomposing VFI into codec-side prealignment (CPU/GPU) and NPU-side residual refinement eliminates the dominant deployment blockers, producing a three-processor pipeline (CPU, GPU, NPU) whose inference graph is composed almost entirely of compute-bound operators. The resulting quality--deployability tradeoff is quantified explicitly (Sec.~\ref{sec:npu_deploy},~\ref{sec:comparison}).
  \item \emph{System validation.} We demonstrate sustained end-to-end 1080p 30$\to$60\,fps playback in an open-source mpv-android fork on SM8650, with 28.4\,ms median VFI latency over 54,623 consecutively logged frame pairs during a 30-minute run (94.9\% within the 33.3\,ms budget), with preliminary cross-vendor feasibility checks on two MediaTek APU generations under the public NeuroPilot SDK (Sec.~\ref{sec:e2e}).
\end{enumerate}

\textbf{Code availability.} Training code, evaluation scripts, and table-reproduction recipes are publicly available at \url{https://github.com/NihilDigit/anvil}; pre-trained checkpoints are hosted on HuggingFace (linked from the repository). Some tables additionally require external vendor repositories (RIFE, IFRNet) or device-side tools (QAIRT SDK, Android device). A separate open-source Android demo is available at \url{https://github.com/NihilDigit/mpv-android-anvil}.

\textbf{Scope.} This work is a deployment-oriented systems study targeting H.264 playback scenarios where codec parameters can be controlled: \texttt{bframes=0}, software decoding with MV side-data export (\texttt{export\_mvs}), and known reference distance. Quality metrics are reported as the measured cost of reaching a deployable operating point under these constraints plus operator compatibility, W8A8 quantization, and sustained thermal requirements. Hardware decoding via \texttt{MediaCodec} does not expose per-macroblock MVs; extending to hardware decode and uncontrolled content is discussed in Sec.~\ref{sec:discussion}.

\section{Related Work}
\label{sec:related}

\subsection{Flow-Based VFI and the NPU Deployment Gap}

Flow-based VFI methods estimate bidirectional optical flow and synthesize intermediate frames via differentiable warping. From a deployment perspective, these methods share three operator-level patterns: (i)~\texttt{grid\_sample}-based spatial warping, established by Super SloMo~\cite{superslomo2018} and adopted by nearly all subsequent flow-based methods; (ii)~iterative or multi-scale flow refinement---RIFE~\cite{rife2022} refines intermediate flow across scales, IFRNet~\cite{ifrnet2022} jointly refines features and flow, and UPR-Net~\cite{uprnet2023} uses a pyramid recurrent design; (iii)~all-pairs attention or multi-field warping---AMT~\cite{amt2023} computes dense correlation volumes, VFIformer~\cite{vfiformer2022} and EMA-VFI~\cite{emavfi2023} introduce transformer or inter-frame attention. Kernel-based methods~\cite{sepconv2017} fold motion estimation into adaptive convolution kernels, while depth-aware methods~\cite{dain2019} add geometric cues. From a mobile-NPU deployment perspective, however, these methods frequently rely on operators such as \texttt{grid\_sample}, multi-scale \texttt{Resize}, and iterative accumulation, which later emerge as the dominant practical barriers under public deployment stacks (Sec.~\ref{sec:int8}).

Network-compression-driven design has been explored for lightweight VFI: CDFI~\cite{cdfi2021} uses network pruning to reduce model complexity for resource-constrained settings. For mobile video deployment more broadly, MobileNVC~\cite{mobilenvc2024} demonstrates real-time 1080p neural video compression on a mobile device through integer quantization and multi-processor co-design, establishing that mobile video models must be architected around operator--accelerator affinity.

Recent work on mobile-efficient architectures reinforces that on-device latency depends on this affinity rather than FLOPs: MobileOne~\cite{mobileone2023} reports phone-measured latency, and EfficientViT~\cite{efficientvit2023} identifies memory-bound operators as the primary speed bottleneck. However, these insights have been developed primarily for classification, detection, and compression; systematic operator-level deployment evidence for mainstream VFI pipelines on public mobile NPU stacks---where \texttt{grid\_sample}, iterative accumulation, and multi-scale \texttt{Resize} dominate the operator mix---remains absent.

\subsection{Codec Motion Priors}

Video codecs compute block-level motion vectors as part of the compression pipeline. The H.264~\cite{h264standard}, HEVC~\cite{hevcstandard2012}, and VVC~\cite{vvcstandard2021} standards provide progressively richer motion models, and early work showed that codec-produced MVs can approximate optical flow~\cite{mpegflow2005} despite their block-level granularity.

Traditional motion-compensated frame interpolation (MCFI) methods directly use codec or block-matching MVs for frame-rate up-conversion: bilateral motion estimation with adaptive OBMC~\cite{bilateralobmc2007}, its triple-frame extension~\cite{tripleframe2019}, and weighted convolutional motion-compensated interpolation~\cite{weightedconv2020}. These signal-processing approaches demonstrate that alignment quality and post-warp refinement are the dominant factors in interpolation quality, but they operate at block level without neural refinement and are limited by block artifacts and occlusion handling.

In the neural domain, codec-side motion has been exploited across several tasks. CoViAR~\cite{coviar2018} uses compressed-domain motion and residuals for action recognition; MVFlow~\cite{mvflow2023} uses codec MVs as a prior for optical flow estimation; CIAF~\cite{ciaf2022} uses codec MVs as an alignment prior for compressed video \emph{super-resolution} (not interpolation), achieving quality comparable to optical-flow alignment at lower cost. Related work also inserts learned prediction inside the codec loop to improve \emph{encoding} efficiency~\cite{deepframe2020,nninterpred2022}, a goal distinct from playback-time interpolation. Most recently, Hint-Guided VFI~\cite{hintvfi2025} exploits compressed-domain hints (low-resolution encoded target frames) to guide neural frame interpolation within a compression pipeline; its hints are pixel-domain representations of the target frame rather than motion-domain priors. We note that these codec-aware neural methods are distinct from VFI architectures that predict learned motion offsets internally (\eg, deformable convolutions); the latter do not reuse decoder-side motion information.

These works collectively show that codec-domain motion cues are useful across compressed-video tasks, and recent work has begun to exploit compressed-domain hints for VFI itself. Our focus differs in targeting \emph{playback-time client-side} interpolation, where decoder-exposed block-level MVs serve as the \emph{primary} motion prior---replacing learned optical flow entirely---and the neural component is co-designed for INT8 execution on public mobile NPU deployment stacks. To our knowledge, this combination of codec MV prealignment as sole motion source with NPU-friendly residual refinement and operator-constrained architecture has not been studied.

\subsection{Deploy-Time Optimization}

Post-training quantization (PTQ)~\cite{jacob2018quantization,gholami2022survey} is the standard route to low-precision NPU deployment. For single-pass convolutional architectures, W8A8 PTQ typically incurs negligible quality loss~\cite{nagel2021whitepaper}. However, quantization becomes harder when activation distributions are shaped by normalization or recurrent refinement; RepQ-ViT~\cite{repqvit2023} showed this clearly for transformer activations. In VFI specifically, the quantization sensitivity of iterative flow refinement under public mobile W8A8 deployment has not been characterized at the operator level; our per-operator causal analysis (Sec.~\ref{sec:int8}) addresses this gap by identifying the specific mechanism---quantized \texttt{Add} on recurrent flow states---through which INT8 error amplifies across refinement stages.

Neither line of work addresses \texttt{grid\_sample} NPU compatibility, iterative flow quantization sensitivity, or the compute-bound operator ratio governing real-time 1080p feasibility.

\section{Proposed Method}
\label{sec:method}

\subsection{System Overview}

\begin{figure*}[!htbp]
\centering
\includegraphics[width=\textwidth]{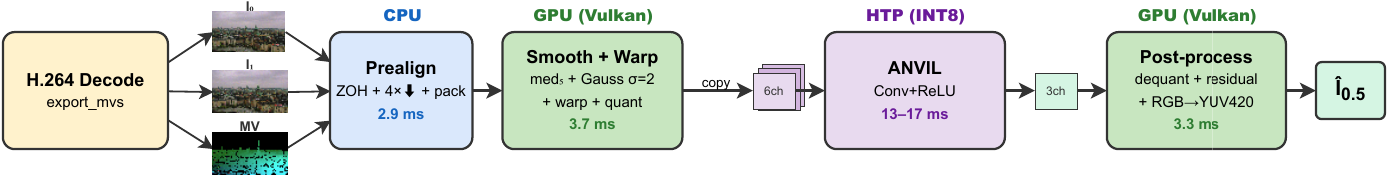}
\caption{ANVIL three-processor pipeline. CPU densifies and downsamples MVs (${\sim}$2.9\,ms); GPU (Vulkan) performs median filtering, Gaussian blur, and sub-pixel remap (${\sim}$3.7\,ms); HTP runs the INT8 residual network (${\sim}$13--17\,ms). CPU/GPU preparation for frame $N{+}1$ is pipelined with HTP inference for frame $N$.}
\label{fig:pipeline}
\end{figure*}

ANVIL decomposes video frame interpolation into two stages (Fig.~\ref{fig:pipeline}):

\begin{enumerate}
  \item \textbf{Codec MV prealignment} (Sec.~\ref{sec:prealign}): Block-level H.264 motion vectors are extracted during decoding, converted to a dense flow field, spatially smoothed, and used to warp both input frames toward the target intermediate time step. This produces a prealigned frame pair where corresponding regions are roughly registered, reducing residual motion to small local displacements.

  \item \textbf{NPU-native residual refinement} (Sec.~\ref{sec:architecture}): An NPU-friendly U-Net, dominated by standard convolutions and simple pointwise operations, takes the 6-channel prealigned input (concatenated warped frames) and predicts a 3-channel RGB residual that is added to the pixel-wise average of the prealigned frames. This motion-compensation-then-residual-refinement decomposition follows a principle established in earlier frame-rate up-conversion work~\cite{weightedconv2020}, but our network is designed around NPU operator constraints with BN fusion at deploy time (Sec.~\ref{sec:deploy}).
\end{enumerate}

This decomposition exploits the fact that the video codec has \emph{already} solved a coarse version of the motion estimation problem. By reusing this information, we avoid learned optical flow entirely, eliminating both \texttt{grid\_sample} from the NPU inference graph and the iterative flow refinement that causes INT8 quantization collapse.

\subsection{Codec MV Extraction and Prealignment}
\label{sec:prealign}

\subsubsection{H.264 Motion Vector Extraction}
\label{sec:mv_extract}

We extract forward motion vectors from P-frames during H.264 decoding. For training data preparation, each frame triplet $(I_0, I_t, I_1)$ is independently encoded as a two-frame H.264 clip (one I-frame followed by one P-frame) using FFmpeg's \texttt{libx264} with \texttt{bframes=0} and \texttt{-preset medium}. This per-triplet encoding ensures every P-frame MV references a high-quality I-frame rather than a previously compressed P-frame, yielding cleaner motion vectors than extraction from a long-GOP stream. At deployment, MVs are instead extracted from the user's video stream via the decoder's \texttt{export\_mvs} side data, where reference quality varies with encoding settings. We validate in Sec.~\ref{sec:discussion} that for $d_\text{ref}=1$ frames (P-frames referencing the immediately preceding frame), CRF and preset variation have negligible impact ($\pm$0.12\,dB), because the spatial smoothing pipeline (Sec.~\ref{sec:prealign}) absorbs MV quality differences. B-frames with $d_\text{ref}>1$ remain a significant factor ($-1.16\dB$); our deployment handles this via selective interpolation (Sec.~\ref{sec:discussion}).

The raw MVs represent block-level displacements from the current frame $I_1$ to the reference frame $I_0$, reported at quarter-pixel precision (\texttt{motion\_scale}${}=4$). Although block-level MVs are coarser than optical flow, prior work has shown that they provide a reasonable approximation~\cite{mpegflow2005}. We negate these vectors to obtain forward flow $\mathbf{f}_{0 \to 1}$ and assign each vector to the corresponding macroblock region via zero-order hold (ZOH), yielding a block-wise dense flow field.

\subsubsection{Spatial Flow Smoothing}

The raw block-level flow field contains discontinuities at macroblock boundaries and occasional outlier vectors from poor codec matches. Our prealignment pipeline addresses these through spatial smoothing:

\begin{enumerate}
  \item \textbf{Downsample} the ZOH flow field by $4\times$ (nearest-neighbor, preserving blockwise structure).
  \item \textbf{Median filter} ($5 \times 5$ kernel) at quarter resolution to suppress outlier vectors while preserving motion edges.
  \item \textbf{Gaussian blur} ($\sigma = 2.0$) at quarter resolution to smooth transitions between blocks. Combined with bilinear upsampling, this is equivalent to $\sigma \approx 8$ at full resolution.
  \item \textbf{Bilinear upsample} back to full resolution.
  \item \textbf{Sub-pixel remap} using bilinear interpolation to warp $I_0$ by $+\mathbf{f}/2$ and $I_1$ by $-\mathbf{f}/2$, producing prealigned frames at the temporal midpoint $t = 0.5$. (Offline evaluation uses OpenCV \texttt{remap}; the deployment pipeline replaces this with a Vulkan compute shader, Sec.~\ref{sec:e2e}.)
\end{enumerate}

Our ablation study (Sec.~\ref{sec:ablation}) confirms that \emph{flow field smoothness} is the dominant factor for prealignment quality: sub-pixel remap alone (no smoothing) scores 29.63\dB, while median$+$Gaussian smoothing reaches 31.20\dB ($+1.57\dB$); cosine-windowed OBMC~\cite{bilateralobmc2007} without smoothing yields 29.58\dB---no better than plain ZOH (Table~\ref{tab:prealign_ablation}). We adopt ZOH$+$spatial smoothing for simplicity and deployment efficiency. The pipeline operates at quarter resolution for computational efficiency. On mobile, CPU handles ZOH densification and $4{\times}$ downsampling ($\sim$2.9\ms), while GPU (Vulkan compute) performs median filtering, Gaussian blur, and sub-pixel remap ($\sim$3.7\ms), for a total prealignment latency under 7\ms (measured in the end-to-end system, Sec.~\ref{sec:e2e}).

\textbf{Zero-parameter MCFI baseline.} Simply averaging the prealigned frames without any neural network (``MV Blend'') achieves 31.20\dB on Vimeo90K ($+5.59\dB$ over naive averaging). The neural residual adds $+2.25\dB$ (ANVIL-S) on top of this strong classical starting point.

\subsection{NPU-Native Residual Architecture}
\label{sec:architecture}

\subsubsection{Design Principles from INT8 Profiling}

Our architecture design is guided by per-operator INT8 profiling on the Hexagon V75 HTP, revealing several non-obvious constraints:

\begin{itemize}
  \item \textbf{Standard Conv 3$\times$3 + ReLU achieves the highest INT8 utilization.} These compute-bound operations achieve 3.2--5.2$\times$ INT8/FP16 speedup on the HTP. In a pure-Conv model, convolution accounts for ${\sim}$59\% of INT8 inference cycles---the only operator class that scales with INT8 SIMD width.
  \item \textbf{DWConv and channel attention degrade INT8 efficiency.} Replacing standard convolutions with depthwise separable convolutions and squeeze-channel attention (SCA) reduces whole-model INT8/FP16 speedup from 3.2--5.2$\times$ to 2.4--2.7$\times$. Per-operator INT8 profiling attributes this to the memory-bound SCA path: Mul$+$GAP consumes 30\% of cycles and residual Add another 17\%, neither benefiting from reduced arithmetic precision.
  \item \textbf{LayerNorm and GELU incur disproportionate cost or lack hardware support.} In a NAFNet~\cite{nafnet2022} HTP FP16 profile, LayerNorm accounts for 57\% of inference cycles versus 19\% for all convolutions combined. On MediaTek APU, LayerNorm and PReLU fail to map to the neural accelerator entirely (NEURON\_UNMAPPABLE); GELU executes at 5.8$\times$ the normalized latency of Conv 3$\times$3.
  \item \textbf{Activation memory dominates at full resolution.} At 1080p, doubling channel width across all encoder--decoder stages adds only 15\% INT8 latency despite 4.4$\times$ more parameters, because activation tensor transfer---not weight computation---is the throughput bottleneck.
\end{itemize}

\subsubsection{UNet-v3b Architecture}

Based on these insights, we design a 4-level asymmetric U-Net~\cite{unet2015} (Fig.~\ref{fig:arch}) restricted to Conv~3$\times$3, ReLU, stride-2 transposed convolution, element-wise Add, and a 1$\times$1 output head:

\begin{figure*}[!htbp]
\centering
\includegraphics[width=\textwidth]{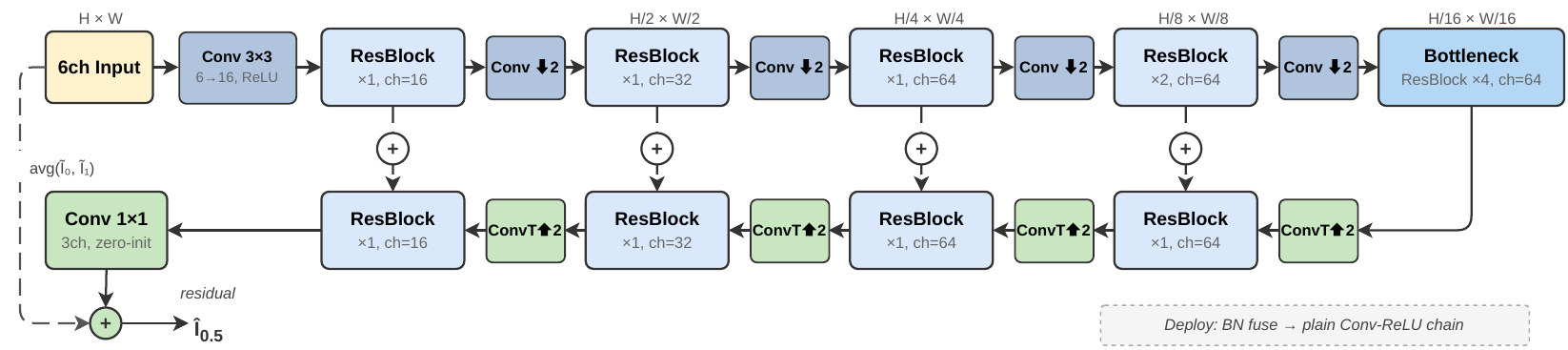}
\caption{UNet-v3b architecture. Input: 6-channel prealigned pair. Encoder: 4 levels with channel widths [16, 32, 64, 64] (ANVIL-S) or [16, 32, 96, 96] (ANVIL-M); encoder blocks per level $(1,1,1,2)$ (ANVIL-S) or $(1,1,2,2)$ (ANVIL-M). Bottleneck: 4 ResBlocks (ANVIL-S) or 8 (ANVIL-M) at $\frac{H}{16} \times \frac{W}{16}$. Decoder uses Add skip connections with 1 ResBlock per level. Output: 3-channel residual added to the prealigned blend. BN is folded into Conv at deploy time.}
\label{fig:arch}
\end{figure*}

\begin{itemize}
  \item \textbf{Asymmetric channels:} Full-resolution stages use narrow channels (16) to limit memory bandwidth, while deep stages use wide channels (64--96) where spatial dimensions are reduced by $8\times$ and compute cost is low.
  \item \textbf{ResBlock with skip connection:} Each block consists of Conv-BN-ReLU-Conv-BN with an element-wise Add skip. BN provides training stability and is fused into Conv weights at deploy time.
  \item \textbf{Zero-initialized output:} The final 1$\times$1 convolution is initialized with zero weights and bias, ensuring the network's initial prediction is the prealigned blend itself. Training then learns to correct residual errors.
  \item \textbf{Element-wise Add skip connections} instead of channel-wise Concat at decoder stages, halving feature map memory at connection points.
\end{itemize}

We provide two configurations:

\begin{center}
\begin{tabular}{lcrr}
\toprule
Variant & Channels & Params & INT8 1080p \\
\midrule
ANVIL-S & $[16, 32, 64, 64]$  & 855K  & 12.8\ms \\
ANVIL-M & $[16, 32, 96, 96]$  & 2.66M & 16.7\ms \\
\bottomrule
\end{tabular}
\end{center}

\subsection{Deploy-Time Optimization}
\label{sec:deploy}

We apply two optimizations to maximize NPU efficiency:

\textbf{(1)~Additive U-Net skip connections.} Decoder skip connections use element-wise Add instead of channel concatenation, eliminating the Concat operations and their associated 1$\times$1 projection convolutions. This choice, guided by INT8 per-operator profiling, reduces total graph operations from 92 to 84 and removes all memory-copy-bound Concat nodes.

\textbf{(2)~BN Fusion.} Batch normalization layers~\cite{batchnorm2015} used during training are folded into preceding convolution weights at deploy time: $W_{\text{fused}} = \frac{\gamma}{\sqrt{\sigma^2 + \epsilon}} W$, $b_{\text{fused}} = \frac{\gamma}{\sqrt{\sigma^2 + \epsilon}}(b - \mu) + \beta$. This is a mathematically exact transformation (maximum absolute difference $< 1.5 \times 10^{-5}$ across 10 random inputs, limited by FP32 rounding from multi-step weight arithmetic) that eliminates all normalization operations from the inference graph while preserving training-time BN benefits.

Together, these optimizations reduce INT8 1080p latency by 17--26\% relative to the baseline concatenation-based U-Net in a same-session A/B test (ANVIL-S: 17.2 $\to$ 12.8\ms; ANVIL-M: 20.2 $\to$ 16.7\ms).

\section{Experiments}
\label{sec:experiments}

\subsection{Experimental Setup}

We organize the evaluation around three questions: (Q1)~What prevents existing VFI from meeting mobile real-time constraints? (Q2)~Does ANVIL remove these blockers under public NPU stacks? (Q3)~What quality is retained? Sec.~\ref{sec:npu_deploy}--\ref{sec:int8} address Q1--Q2; Sec.~\ref{sec:comparison} addresses Q3.

\textbf{Datasets.} We train on Vimeo90K~\cite{toflow2019} (51,313 training triplets) and evaluate on both the Vimeo90K test set (3,782 triplets) and Xiph 1080p~\cite{xiph} (2,662 triplets from 12 sequences at $1080 \times 1920$, constructed as stride-1 consecutive frame triplets---frames $2k$, $2k{+}1$, $2k{+}2$---from each sequence). The Xiph benchmark tests generalization to high-resolution real-world video with diverse motion characteristics.

\textbf{Metrics.} We report PSNR and SSIM~\cite{ssim2004} computed on RGB uint8 images, and LPIPS~\cite{lpips2018} (AlexNet backbone) as a perceptual quality metric.

\textbf{Training.} Models are trained with L1 loss only (no perceptual loss), AdamW optimizer ($\text{lr} = 2 \times 10^{-4}$), batch size 16, AMP bfloat16, and \texttt{torch.compile}. Early stopping with patience 7 on validation PSNR (interval 3 epochs, minimum delta 0.10\dB). Data augmentation includes random crops ($256 \times 256$), horizontal/vertical flips, and temporal reversal. All ANVIL models use prealigned input.

\textbf{Baselines.} Our primary external baseline is RIFE HDv3~\cite{rife2022} (3.04M parameters). We evaluate RIFE at native resolution and with resolution reduction strategies (360p/480p flow-upsample and frame-upsample).

\textbf{Hardware.} Our primary latency and end-to-end measurements are collected on three Qualcomm devices, with separate cross-vendor deployment validation on two MediaTek devices under the public NeuroPilot toolchain:
\begin{itemize}
  \item \textbf{Qualcomm:} Snapdragon 7+ Gen~2 (HTP V69), 8 Gen~2 (V73), 8 Gen~3 (V75), deployed via QNN SDK~\cite{qualcomm_qnn}.
  \item \textbf{MediaTek:} Dimensity 9300 (APU~790), Dimensity 9400+ (APU~890), deployed via NeuroPilot Public SDK~\cite{mediatek_neuropilot}. On this path we validated operator support, full-model 1080p INT8 execution, and small-set on-device quality sanity checks. Because MediaTek uses public-SDK runtime compilation rather than Qualcomm-style offline-compiled contexts, we do not place the absolute latency numbers in the same table as the Qualcomm results. Note that MediaTek's premium SDK (requiring NDA) may support additional operators; our conclusions here are restricted to the publicly available toolchain.
\end{itemize}

\textbf{Quantization protocol.} All INT8 results use static W8A8 post-training quantization with percentile 99.99 calibration. \emph{Calibration data} is drawn from a distribution matching each model's deployment scenario: for ANVIL, 100 samples stratified by motion magnitude from the Vimeo90K \emph{training} set (upscaled to 1080p), since the model operates on prealigned training-distribution inputs; for cross-method baselines (RIFE, IFRNet), calibration samples are drawn from Xiph 1080p frames downsampled to each model's input resolution (360p or 480p), matching the deployment pipeline of reduced-resolution inference on high-resolution content. Calibration and evaluation are split at the \emph{sequence level}: two Xiph sequences (\texttt{sunflower}, \texttt{pedestrian\_area}) are reserved exclusively for calibration; evaluation uses only the remaining ten sequences, ensuring zero content overlap. For Qualcomm HTP: QNN SDK (QAIRT 2.42), per-tensor symmetric activation quantization, models compiled offline to HTP context binaries. For ONNX Runtime: \texttt{quantize\_static} with QOperator format, per-tensor activation quantization. For MediaTek: TFLite INT8 via \texttt{mtk\_converter} with default symmetric quantization. All latency measurements use BURST profile, 50--100 iterations after 10 warm-up, reporting minimum latency to exclude scheduling jitter.

\subsection{NPU Deployment Analysis}
\label{sec:npu_deploy}

We first establish why existing VFI methods cannot be deployed at 1080p INT8 on current mobile NPUs, before evaluating ANVIL's quality under these constraints.

\subsubsection{Cross-Device Latency}

Table~\ref{tab:latency} presents INT8 1080p latency across three Qualcomm generations. ANVIL-S meets the 33.3\ms deadline on V73/V75; on V69, 720p remains viable (10.5\ms).

\begin{table}[!htbp]
\centering
\caption{HTP INT8 network inference latency at 1080p (NPU forward pass only; end-to-end in Table~\ref{tab:e2e_latency}). $^\dagger$720p fallback: ANVIL-S 10.5\ms, ANVIL-M 14.1\ms. $^\ddagger$RIFE at 360p.}
\label{tab:latency}
\renewcommand{\arraystretch}{1.15}
\resizebox{\columnwidth}{!}{%
\begin{tabular}{l c r r r c}
\toprule
SoC & NPU & ANVIL-S & ANVIL-M & RIFE$^\ddagger$ & $\leq$33.3ms \\
\midrule
SD 7+ Gen2 & HTP V69  & 46.0\ms & 56.6\ms & 45.6\ms & $\times^\dagger$ \\
SD 8 Gen2  & HTP V73  & 15.5\ms & 20.8\ms & 20.7\ms & \checkmark \\
SD 8 Gen3  & HTP V75  & 12.8\ms & 16.7\ms & 14.7\ms & \checkmark \\
\bottomrule
\end{tabular}}
\end{table}

\textbf{ANVIL-S vs.\ RIFE latency.} On the HTP V75, ANVIL-S at full 1080p INT8 (12.8\ms) runs faster than RIFE at 360p INT8 (14.7\ms), despite processing $9\times$ more pixels. This reflects ANVIL's compute-bound graph versus RIFE's 5.1\% Conv ratio (95\% memory-bound operations including Resize, GridSample, and element-wise arithmetic).

\textbf{FP16 is insufficient.} All models---including RIFE at 360p---exceed the 33.3\ms deadline under FP16 on all tested devices. Under the studied public deployment stacks, W8A8 is the practical operating point for 1080p mobile VFI.

\textbf{MediaTek validation.} Under NeuroPilot Public SDK, ANVIL-S executes at 1080p INT8 on both tested generations (24.4\ms on D9300, 25.5\ms on D9400+) with all operators APU-delegated and negligible INT8 loss. These are cross-vendor deployment validations rather than directly comparable latency points, as the public MediaTek path uses runtime compilation whose overhead dominates absolute timing.

\textbf{Deployment audit of existing methods.} To confirm that the deployment barriers are not RIFE-specific, we attempted full deployment pipelines for IFRNet~\cite{ifrnet2022} (5.0M, 4-level iterative refinement) and AMT-S~\cite{amt2023} (3.0M, 3-stage correlation + 32$\times$\texttt{GridSample}). IFRNet exports and compiles successfully in our audit, but the archived artifacts do not establish a quality-preserving 1080p real-time operating point; Sec.~\ref{sec:int8} shows that under a unified PTQ protocol its frame-output path loses 2.95\,dB and even the flow-up path still requires a full-resolution warp stage. AMT-S fails HTP context binary generation entirely (``Graph Finalize failure''). These failures are structural, not implementation-specific.

\subsubsection{Cross-Vendor Operator Compatibility}

\begin{table*}[!htbp]
\centering
\caption{Cross-vendor operator compatibility for 17 VFI operators at 1080p. Latency normalized to Conv 3${\times}$3 = 1.0${\times}$. ``Freq'': methods using the operator (out of 9 surveyed). Groups: (A)~universal, (B)~partial, (C)~limited, (D)~iterative patterns.}
\label{tab:operator_compat}
\renewcommand{\arraystretch}{1.12}
\begin{tabular}{c l c c c l}
\toprule
Group & Operator & Freq & HTP V75 & APU 790 & Impact \\
\midrule
A & Conv2d 3$\times$3        & 9/9 & 1.0$\times$  & 1.0$\times$ & Baseline \\
A & Conv2d 1$\times$1        & 8/9 & 0.9$\times$  & 1.0$\times$ & \\
A & Conv + ReLU              & 5/9 & 1.0$\times$  & 1.0$\times$ & \\
A & ConvTranspose2d          & 7/9 & 3.9$\times$  & 2.7$\times$ & Moderate cost both platforms \\
A & Residual Add             & 9/9 & 2.3$\times$  & 1.8$\times$ & \\
\midrule
B & \texttt{GridSample}      & 7/9 & 3.2$\times$  & \textbf{N/A}   & MTKEXT custom op, public SDK unavailable \\
B & Resize 2$\times$         & 7/9 & 4.4$\times$  & \textbf{12.6$\times$} & Both platforms bottleneck \\
B & Conv + Sigmoid           & 7/9 & 1.1$\times$  & 0.9$\times$ & \\
B & Conv + PReLU             & 1/9 & 1.0$\times$  & \textbf{UNMAP} & RIFE blocked on MediaTek \\
B & Conv + LeakyReLU         & 2/9 & 1.0$\times$  & 1.0$\times$ & \\
\midrule
C & Conv + LayerNorm         & 2/9 & 1.5$\times$  & \textbf{UNMAP} & Transformer VFI blocked on MTK \\
C & Conv + GELU              & 1/9 & 1.1$\times$  & \textbf{5.8$\times$} & \\
C & DWConv 3$\times$3        & 1/9 & 1.0$\times$  & 0.9$\times$ & \\
C & Self-Attention           & 2/9 & OOM          & OOM         & 1080p not exportable \\
C & Deformable Conv          & 1/9 & N/A          & N/A         & ONNX unsupported \\
\midrule
D & \texttt{iter\_accum\_3}  & 4/9 & \textbf{23.8$\times$} & 1.7$\times$ & HTP 23.8$\times$ overhead; INT8 collapses \\
D & \texttt{warp\_chain}     & 7/9 & 5.8$\times$  & \textbf{N/A}   & Contains GridSample \\
\bottomrule
\end{tabular}
\end{table*}

Table~\ref{tab:operator_compat} presents a systematic operator-level benchmark across both NPU vendors, testing 17 operators drawn from 9 prominent VFI methods~\cite{rife2022,ifrnet2022,amt2023,vfiformer2022,emavfi2023,m2m2022,abme2021,flavr2023,uprnet2023}. The results reveal that \textbf{operators used by ANVIL (standard convolutions and simple pointwise operations) are the only ones achieving $\sim$1.0$\times$ baseline performance on both platforms}. In contrast:

\begin{itemize}
  \item \texttt{GridSample}, used by 7/9 methods, is mapped to a vendor-specific custom op (\texttt{MTKEXT\_GRID\_SAMPLE\_2D}) unavailable through MediaTek's public NeuroPilot SDK (the premium SDK with NDA access may support it), and incurs 3.2$\times$ overhead on Qualcomm HTP.
  \item PReLU and LayerNorm are unmappable on MediaTek APU entirely, blocking RIFE and transformer-based VFI.
  \item Resize 2$\times$ is a bottleneck on both platforms (4.4$\times$ HTP, 12.6$\times$ APU), penalizing multi-scale pyramid approaches.
  \item Self-Attention causes OOM at 1080p on both platforms.
\end{itemize}

\textbf{RIFE HTP per-operator profile.} Profiling RIFE 360p FP16 on HTP V75 confirms the structural inefficiency: convolutions account for only 5.1\% of inference cycles, while memory-bound operations consume 95\% (\texttt{Resize}: 26.6\%, \texttt{GridSample}: 15.6\%, \texttt{Div}: 13.6\%, \texttt{Add}: 8.9\%, \texttt{Mul}: 8.4\%, \texttt{Concat}: 6.9\%, \texttt{PReLU}: 6.5\%, \texttt{Slice}: 4.6\%). By contrast, ANVIL's inference graph ensures that the majority of NPU cycles are spent on compute-bound operations.

The operator compatibility and latency data establish that no tested VFI method simultaneously meets the 33.3\ms latency budget and the cross-vendor operator constraint. We next evaluate whether INT8 quantization preserves quality.

\subsection{INT8 Quantization Analysis}
\label{sec:int8}

The preceding section established that INT8 is the only precision mode meeting latency budgets. We now evaluate whether INT8 quantization preserves interpolation quality, comparing ANVIL, RIFE~\cite{rife2022}, and IFRNet~\cite{ifrnet2022} using two backends overall---QNN on-device for ANVIL (matching the deployment target) and ORT CPU for RIFE/IFRNet (reproducible offline baseline)---to distinguish architectural quantization sensitivity from backend-specific effects. AMT-S~\cite{amt2023} cannot be included because its 32 \texttt{GridSample} operators cause HTP context compilation to fail.

\subsubsection{Cross-Method INT8 Quality}

\begin{table}[!htbp]
\centering
\caption{INT8 quantization quality on Xiph 1080p. ``Mode'': flow$\uparrow$ outputs flow for CPU warp; frame outputs RGB directly. $^\dagger$Full 2,662-triplet QNN on-device evaluation (aggregate PSNR; per-sample bootstrap CI not available for device-side inference). Calibration and eval protocol details in Sec.~\ref{sec:experiments}.}
\label{tab:int8_quality}
\renewcommand{\arraystretch}{1.15}
\resizebox{\columnwidth}{!}{%
\begin{tabular}{l c l c l}
\toprule
Method & Stg & Mode & PSNR: FP32\,$\to$\,INT8 ($\Delta$) & 95\% CI \\
\midrule
\textbf{ANVIL-S} & \textbf{0} & \textbf{direct} & \textbf{29.28}$\,\to\,$\textbf{29.09} ($\boldsymbol{-}$\textbf{0.19})$^\dagger$ & ---$^\dagger$ \\
\textbf{ANVIL-M} & \textbf{0} & \textbf{direct} & \textbf{29.37}$\,\to\,$\textbf{29.28} ($\boldsymbol{-}$\textbf{0.09})$^\dagger$ & ---$^\dagger$ \\
RIFE   & 3 & flow$\uparrow$ & 26.79$\,\to\,$25.90 ($-$0.89) & [$-$0.96, $-$0.83] \\
IFRNet & 4 & flow$\uparrow$ & 25.43$\,\to\,$25.03 ($-$0.40) & [$-$0.44, $-$0.37] \\
IFRNet & 4 & frame          & 25.84$\,\to\,$21.46 ($-$4.38) & [$-$4.75, $-$4.01] \\
\bottomrule
\end{tabular}}
\end{table}

Table~\ref{tab:int8_quality} presents INT8 quality on Xiph 1080p under the best archived protocols available for each method. Three patterns emerge. First, \textbf{ANVIL shows negligible degradation}: ANVIL-S loses $-0.19\dB$ and ANVIL-M $-0.09\dB$ in the archived full 2,662-triplet QNN on-device evaluation. Second, \textbf{iterative flow methods degrade}, with severity depending on mode: IFRNet frame mode collapses by $-4.38\dB$ (CI [$-4.75$, $-4.01$], 60\% of samples $>$3\dB), while flow-up modes show moderate degradation (RIFE $-0.89\dB$, IFRNet $-0.40\dB$). All confidence intervals exclude zero. Third, \textbf{flow-up mitigates but does not eliminate} the problem---and still requires a full-resolution warp stage outside the low-resolution network.

\subsubsection{Per-Operator Causal Analysis}

To identify the mechanism, we progressively add operator types to ORT W8A8 quantization and measure output cosine similarity (CosSim) against the FP32 baseline. We perform this analysis on both RIFE and IFRNet with trained weights.

\begin{table}[!htbp]
\centering
\caption{Per-operator instrumented INT8 quantization. CosSim measured against FP32 output. The same causal pattern holds for both methods: Conv introduces initial error, PReLU has zero effect, and \textbf{Add triggers the collapse}. Full W8A8 $\approx$ Conv$+$Add.}
\label{tab:instrumented}
\renewcommand{\arraystretch}{1.15}
\begin{tabular}{l c c c c}
\toprule
\multirow{2}{*}{Quantized Ops} & \multicolumn{2}{c}{RIFE (3-stage)} & \multicolumn{2}{c}{IFRNet (4-stage)} \\
\cmidrule(lr){2-3} \cmidrule(lr){4-5}
 & CosSim & $\Delta$ & CosSim & $\Delta$ \\
\midrule
None (FP32)        & 1.000 & ---      & 1.000 & ---      \\
Conv, ConvTr       & 0.952 & $-$0.048 & 0.945 & $-$0.055 \\
$+$ PReLU          & 0.952 & 0.000    & 0.945 & 0.000    \\
\textbf{$+$ Add}   & \textbf{0.815} & $-$\textbf{0.137} & \textbf{0.878} & $-$\textbf{0.068} \\
Full W8A8          & 0.790 & $\approx$Add & 0.859 & $\approx$Add \\
Resize only        & 1.000 & 0.000    & 1.000 & 0.000    \\
\bottomrule
\end{tabular}
\end{table}

The causal chain (Table~\ref{tab:instrumented}) is identical for both methods:
\begin{enumerate}
  \item \textbf{Conv quantization} introduces initial error ($-0.048$ for RIFE, $-0.055$ for IFRNet), acceptable in isolation.
  \item \textbf{PReLU quantization} has zero additional effect in both methods, confirming it is benign.
  \item \textbf{Add quantization triggers collapse.} Add implements iterative flow accumulation ($\mathbf{f}_{\text{accum}} {=} \mathbf{f}_{\text{accum}} {+} \Delta\mathbf{f}$); each stage's quantized output feeds the next, amplifying error across iterations.
  \item \textbf{Full W8A8 $\approx$ Conv $+$ Add}---no other operator contributes.
\end{enumerate}

Notably, individual operators are robust in isolation: a standalone 3-stage iterative Add with random data achieves CosSim $= 0.9999$. The collapse emerges only when \emph{trained weights} produce flow states with large dynamic range ($\pm 19$ pixels for RIFE, $\pm 11$ for IFRNet) that are iteratively accumulated through quantized Add operations.

\subsubsection{Why ANVIL Is Immune}

ANVIL's architecture avoids all three quantization risk factors identified above:
\begin{itemize}
  \item \textbf{No iterative accumulation.} Adds are single-pass residual skips ($\text{output} = \text{blend} + \text{residual}$), no recurrent state.
  \item \textbf{Small dynamic range.} Residual range $\pm 0.25$ (vs.\ $\pm 19$ px for RIFE flow), well within 8-bit capacity.
  \item \textbf{No \texttt{grid\_sample}.} Warping occurs outside the NPU graph, so sampling does not amplify quantization noise.
\end{itemize}
These structural differences explain the large gap in INT8 degradation: ANVIL loses only $-0.19\dB$, while iterative flow systems degrade more strongly ($-0.89\dB$ for RIFE flow$\uparrow$, $-4.38\dB$ for IFRNet frame mode).

\textbf{Implications for quality evaluation.} The deployment and INT8 analyses above establish that ANVIL is the \emph{only} VFI method in our evaluation with a fully verified 1080p INT8 operating point on Qualcomm HTP and an additional public-SDK deployment validation on two MediaTek APU generations. With this context, we now evaluate interpolation quality.

\subsection{Quality Under Deployment Constraints}
\label{sec:comparison}

\begin{table*}[!htbp]
\centering
\caption{Quality and deployment status on Vimeo90K and Xiph 1080p. \textbf{Bold}: best among deployable methods. For context, non-deployable methods reach 35--36\,dB on Vimeo90K (IFRNet~\cite{ifrnet2022} 35.80, AMT-S~\cite{amt2023} 35.72, EMA-VFI~\cite{emavfi2023} 36.11).}
\label{tab:main_quality}
\renewcommand{\arraystretch}{1.12}
\begin{tabular}{l r c c c c c c l}
\toprule
\multirow{2}{*}{Method} & \multirow{2}{*}{Params} & \multicolumn{3}{c}{Vimeo90K} & \multicolumn{3}{c}{Xiph 1080p} & \multirow{2}{*}{Deployment Status} \\
\cmidrule(lr){3-5} \cmidrule(lr){6-8}
 & & PSNR & SSIM & LPIPS & PSNR & SSIM & LPIPS & \\
\midrule
\multicolumn{9}{l}{\textit{Zero-parameter baselines}} \\
Naive Blend       & 0     & 25.61 & 0.753 & 0.099 & 25.32 & 0.658 & 0.165 & --- \\
MV Blend       & 0     & 31.20 & 0.926 & 0.053 & 28.98 & 0.813 & 0.115 & --- \\
\midrule
\multicolumn{9}{l}{\textit{ANVIL (ours) --- 1080p INT8 deployable}} \\
ANVIL-S           & 855K  & \textbf{33.45} & \textbf{0.949} & 0.037 & \textbf{29.65} & \textbf{0.835} & 0.154 & \textbf{Deployable} \\
ANVIL-M           & 2.66M & \textbf{33.66} & \textbf{0.951} & \textbf{0.036} & \textbf{29.74} & \textbf{0.836} & 0.148 & \textbf{Deployable} \\
\midrule
\multicolumn{9}{l}{\textit{Flow/warp methods --- deployment blocked (Sec.~\ref{sec:npu_deploy},~\ref{sec:int8})}} \\
RIFE~\cite{rife2022}  & 3.04M & 34.26 & 0.956 & 0.019 & 30.04 & 0.828 & 0.077 & grid\_sample + recurrent flow \\
RIFE 360p flow$\uparrow$  & 3.04M & ---   & ---   & ---   & 29.19 & 0.805 & 0.113 & INT8 collapse (Table~\ref{tab:rife_reduced}) \\
RIFE 480p flow$\uparrow$  & 3.04M & ---   & ---   & ---   & 29.77 & 0.820 & 0.103 & INT8 collapse (Table~\ref{tab:rife_reduced}) \\
\midrule
\multicolumn{9}{l}{\textit{Architecture ceiling (not deployable)}} \\
NAFNet-ceiling     & 17.1M & 34.58 & 0.959 & 0.030 & 30.30 & 0.850 & 0.175 & LayerNorm 57\% cycles \\
\bottomrule
\end{tabular}
\end{table*}

Having established in Sec.~\ref{sec:npu_deploy} and~\ref{sec:int8} that no tested iterative flow method meets the 1080p latency budget while surviving INT8 quantization, Table~\ref{tab:main_quality} jointly evaluates interpolation quality alongside deployment status. We benchmark RIFE~\cite{rife2022} under our unified protocol as the primary flow-based reference, including reduced-resolution variants at 360p and 480p with on-device INT8 latency verification---both within the 33.3\ms budget but suffering INT8 quality collapse on two independent backends ($-2.03\dB$ ORT at 360p flow$\uparrow$; Table~\ref{tab:rife_reduced}). The perceptual gap is notable: RIFE achieves substantially better LPIPS (0.019 vs.\ 0.036 on Vimeo, 0.077 vs.\ 0.148 on Xiph), reflecting the smoothness bias inherent in residual prediction versus sub-pixel warping. Among deployable configurations, ANVIL-M achieves the highest quality on both datasets. Traditional MCFI methods~\cite{bilateralobmc2007,tripleframe2019,weightedconv2020} do not provide Vimeo90K or Xiph evaluations under comparable protocols; our MV Blend baseline, which applies the same codec-MV smoothing pipeline without neural refinement, serves as the most direct classical reference point.

\textbf{Deployment--quality tradeoff.} The quality gap between ANVIL-M and non-deployable RIFE native (33.66 vs.\ 34.26\dB on Vimeo, 29.74 vs.\ 30.04\dB on Xiph) reflects the cost of eliminating \texttt{grid\_sample} and iterative flow from the NPU graph. A NAFNet ceiling model (17.1M, same prealigned input) reaches 34.58\dB---still below sub-pixel flow methods (IFRNet 35.80, EMA-VFI 36.11)---suggesting $\sim$0.9\dB from capacity constraints and $\sim$1.2\dB from the structural limitation of residual prediction versus learned warping. The LPIPS gap (${\sim}2{\times}$: 0.036 vs.\ 0.019 on Vimeo; 0.148 vs.\ 0.077 on Xiph) persists even in the NAFNet ceiling, indicating a structural property of the residual paradigm rather than a capacity issue. This is the primary perceptual cost of NPU-native design, one that could be partially mitigated by perceptual loss fine-tuning (though such tuning must be validated against INT8 robustness); temporal quality analysis follows in Sec.~\ref{sec:discussion}.

\begin{figure*}[!htbp]
\centering
\includegraphics[width=0.85\textwidth]{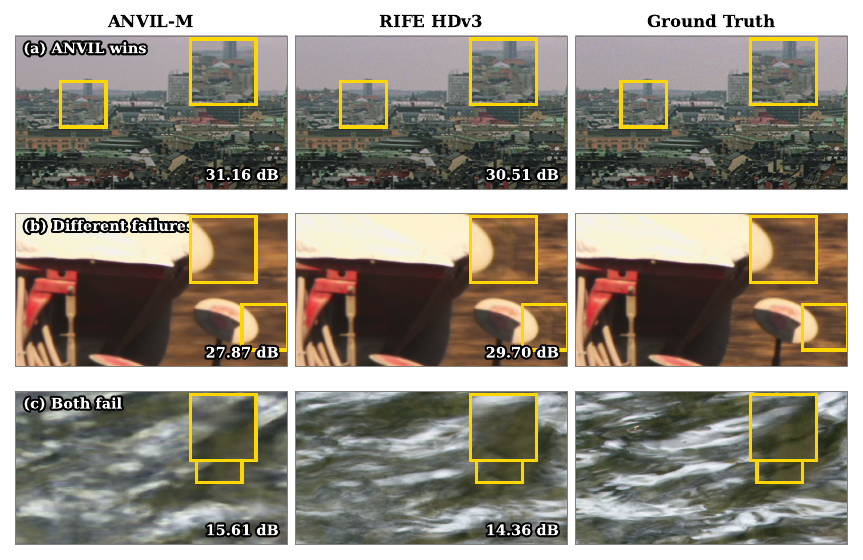}
\caption{Visual comparison on Xiph 1080p ($3\times$ magnified insets). (a)~\texttt{old\_town\_cross}: ANVIL smoothing suppresses noise, RIFE preserves detail. (b)~\texttt{tractor}: ANVIL over-smooths edges, RIFE produces ghosting. (c)~\texttt{riverbed}: both fail on stochastic texture.}
\label{fig:visual}
\end{figure*}

\textbf{Deployment quality analysis.} Fig.~\ref{fig:visual} illustrates representative failure modes. On \texttt{old\_town\_cross} (slow pan), ANVIL's smoothing suppresses noise ($+0.7\dB$ vs.\ RIFE) but loses fine detail. On \texttt{tractor} (large rigid motion), ANVIL over-smooths edges while RIFE produces ghosting; neither is artifact-free. On \texttt{riverbed} (stochastic texture), all methods collapse as non-rigid motion defeats both paradigms. The smoothness bias from spatial flow smoothing benefits low-texture regions but penalizes high-frequency detail, explaining ANVIL's LPIPS disadvantage.

\textbf{Resolution reduction strategies.} Table~\ref{tab:rife_reduced} evaluates RIFE at 360p and 480p in two modes: \emph{flow-upsample} (low-res flow, CPU warp at full resolution) and \emph{frame-upsample} (entire pipeline at low resolution, bicubic output). At FP32, 480p flow-upsample (29.13\dB) approaches ANVIL-M (29.37\dB). However, INT8 collapses on both ORT and QNN backends ($-2.03$ to $-5.12\dB$); frame-upsample degrades more severely because the quantized warp path amplifies errors internally. All FP16 latencies exceed 33.3\ms, and 480p frame-upsample exceeds the deadline even at INT8. No RIFE configuration simultaneously satisfies latency, quality, and INT8 robustness. FP32 values differ from Table~\ref{tab:main_quality} because two sequences are reserved for INT8 calibration (Sec.~\ref{sec:experiments}).

\begin{table}[!htbp]
\centering
\caption{RIFE resolution reduction strategies on Xiph 1080p. ORT $\Delta$: W8A8 via ORT CPU. QNN $\Delta$: W8A8 vs FP16 on HTP V75. Both backends confirm INT8 degradation; all FP16 latencies exceed 33.3\ms.}
\label{tab:rife_reduced}
\renewcommand{\arraystretch}{1.15}
\begin{tabular}{l l r r r r}
\toprule
Res. & Mode & FP32 & ORT $\Delta$ & QNN $\Delta$ & INT8 lat. \\
\midrule
360p & flow$\uparrow$  & 28.54 & $-$2.03 & $-$1.43 & 14.7\ms \\
360p & frame$\uparrow$ & 27.00 & $-$3.80 & $-$3.44 & 17.8\ms \\
480p & flow$\uparrow$  & 29.13 & $-$2.90 & $-$2.24 & 28.2\ms \\
480p & frame$\uparrow$ & 28.23 & $-$5.12 & $-$4.84 & 36.3\ms$^*$ \\
\bottomrule
\multicolumn{6}{l}{\footnotesize $^*$Exceeds 33.3\ms deadline even at INT8.} \\
\end{tabular}
\end{table}

\subsection{Ablation Studies}
\label{sec:ablation}

\subsubsection{Prealignment Contribution}

MV prealignment is the single most impactful design choice. A same-capacity Route~A baseline (A-small, 33K parameters, no prealignment) achieves only 29.00\dB on Vimeo---worse than the zero-parameter MV Blend (31.20\dB, Table~\ref{tab:main_quality}). Replacing the production smoothing pipeline with a legacy block-averaging scheme (v1: box filter $+$ $16{\times}16$ block average) degrades MV Blend from 31.20 to 29.92\dB on Vimeo ($-1.28\dB$) and from 28.98 to 28.26\dB on Xiph 1080p ($-0.72\dB$), with 95.3\% of Vimeo test samples improving under the production pipeline. From a deployment perspective, prealignment offloads motion estimation from the NPU entirely, enabling the NPU-friendly residual architecture.

\begin{table}[!htbp]
\centering
\caption{Prealignment method ablation on Vimeo90K test set (3,782 triplets, MV Blend PSNR). All methods use the same block-level H.264 MVs; they differ in flow construction and post-processing.}
\label{tab:prealign_ablation}
\renewcommand{\arraystretch}{1.15}
\begin{tabular}{l l c}
\toprule
Flow construction & Post-processing & PSNR \\
\midrule
ZOH & none (sub-pixel remap only)          & 29.63 \\
OBMC (cosine) & none                       & 29.58 \\
Daala & Gaussian $\sigma{=}6$              & 30.89 \\
\textbf{ZOH} & \textbf{median 5$+$Gaussian $\boldsymbol{\sigma{\approx}8}$} & \textbf{31.20} \\
\bottomrule
\end{tabular}
\end{table}

\subsubsection{Capacity Scaling}

\begin{table}[!htbp]
\centering
\caption{Capacity scaling under the 33.3\ms latency budget. Prealigned variants (Route~D) are fully trained on Vimeo90K with basic prealignment. A-small (Route~A) receives raw frames without MV prealignment.}
\label{tab:ablation_capacity}
\renewcommand{\arraystretch}{1.15}
\begin{tabular}{l r c c}
\toprule
Model & Params & Vimeo PSNR & INT8 1080p \\
\midrule
\multicolumn{4}{l}{\textit{Route~A (no prealignment)}} \\
A-small        & 33K   & 29.00 & --- \\
\midrule
\multicolumn{4}{l}{\textit{Route~D (MV prealignment)}} \\
D-tiny-nomv    & 1.8K  & 30.67 & 3.52\ms \\
D-mini-nomv    & 10.6K & 31.08 & 4.26\ms \\
D-mid-nomv     & 33K   & 31.38 & 8.16\ms \\
D-unet-s-nomv  & 129K  & 31.83 & 8.84\ms \\
D-unet-l-nomv  & 289K  & 32.49 & 15.5\ms \\
\bottomrule
\end{tabular}
\end{table}

\begin{figure}[!htbp]
\centering
\includegraphics[width=\columnwidth]{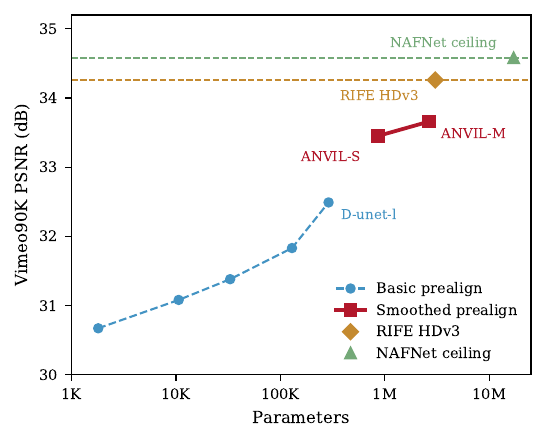}
\caption{Quality scaling with model capacity. Basic prealignment (dashed) and smoothed prealignment (solid) series both scale without saturation. Dashed horizontal lines show RIFE HDv3 (3.04M) and NAFNet ceiling (17.1M, smoothed prealignment retrained) as reference points.}
\label{fig:pareto}
\end{figure}

Table~\ref{tab:ablation_capacity} and Fig.~\ref{fig:pareto} show that quality scales continuously with model capacity, with no saturation across the 1.8K--289K parameter range under basic prealignment. Smoothed prealignment (Sec.~\ref{sec:prealign}) shifts the entire curve upward: ANVIL-S (855K) and ANVIL-M (2.66M) continue the scaling trend toward the NAFNet ceiling (34.58\dB at 17.1M), confirming that the framework has not reached its quality ceiling. The U-Net multi-scale design yields the highest quality-per-latency gain: D-mid $\to$ D-unet-s adds $+0.45\dB$ with only $1.08\times$ latency increase.

\subsubsection{Deploy-Time Optimization}

Replacing concatenation-based U-Net skip connections with additive skips, combined with BN fusion, reduces INT8 1080p latency by 17--26\% in a same-session A/B test (ANVIL-S: 17.2 $\to$ 12.8\ms; ANVIL-M: 20.2 $\to$ 16.7\ms) by eliminating memory-bound Concat and projection operations from the inference graph.

\subsubsection{Final Design}

The final ANVIL-S and ANVIL-M models integrate three improvements over the D-nomv scaling series: (1)~smoothed prealignment (Sec.~\ref{sec:prealign}), which raises the zero-parameter MV~Blend baseline by $+1.28\dB$; (2)~an asymmetric-channel U-Net with narrow full-resolution stages and wide bottleneck stages, designed from INT8 per-operator profiling to maximize compute-bound Conv ratio; and (3)~additive U-Net skip connections with deploy-time BN fusion. Quality and latency are in Table~\ref{tab:main_quality} and Table~\ref{tab:latency}.

\subsection{End-to-End System Validation}
\label{sec:e2e}

We implemented the full ANVIL pipeline as a video filter (\texttt{vf\_anvil}) in an open-source Android player, integrating H.264 software decoding with MV extraction, GPU compute shaders, and HTP INT8 inference on a single SoC. Software decoding (FFmpeg \texttt{libavcodec}) is used because Android's \texttt{MediaCodec} does not expose per-macroblock MVs. Table~\ref{tab:e2e_latency} reports per-stage VFI latency under concurrent software decoding; decode runs on a separate thread but competes for CPU and memory bandwidth, so timing reflects realistic contention.

\subsubsection{Pipeline Architecture}

The pipeline processes each frame pair through five stages distributed across CPU, GPU, and HTP:

\begin{table}[!htbp]
\centering
\caption{End-to-end per-stage latency on SM8650, ANVIL-S 1080p INT8. Medians over 54,623 consecutive frame pairs (30-minute playback).}
\label{tab:e2e_latency}
\renewcommand{\arraystretch}{1.15}
\resizebox{\columnwidth}{!}{%
\begin{tabular}{l l r l}
\toprule
Stage & HW & Latency & Operation \\
\midrule
P1a & CPU    & 2.9\ms & ZOH densify + $4{\times}$ downsample + YUV pack \\
P1b+P2 & GPU & 3.7\ms & median$_{5}$ + Gauss $\sigma\!=\!2$ + warp + quant \\
Copy & CPU    & 0.9\ms & 12\,MB uint8 NHWC to QNN buffer \\
P3  & HTP     & 17.0\ms & INT8 graphExecute (async, pipelined) \\
P4  & GPU     & 3.3\ms & dequant + residual + RGB$\to$YUV420 \\
\midrule
\multicolumn{2}{l}{\textbf{Total per pair}} & \textbf{28.4\ms} & median; 94.9\% $\leq$33.3\ms \\
\bottomrule
\end{tabular}}
\end{table}

Three optimizations were critical for achieving this latency:

\textbf{GPU-side quantization.} A fused Vulkan compute shader performs flow upsampling, warp, blend, and INT8 quantization in a single dispatch, producing a compact uint8 buffer that replaces ${\sim}$71\,MB of intermediate float32 tensors. This reduces the CPU$\to$HTP copy from ${\sim}$8\ms to ${\sim}$0.9\ms.

\textbf{Pipelined HTP with double-buffered I/O.} A dedicated inference thread with double-buffered pending state enables frame-level pipeline overlap: CPU/GPU preparation for frame $N{+}1$ runs concurrently with HTP inference for frame $N$, so HTP and CPU/GPU stages execute in parallel.

\textbf{GPU post-processing.} Dequantization, residual addition, and RGB-to-YUV420 conversion are moved from CPU to a GPU compute shader, reducing this phase from 11--21\ms (CPU, variable with big.LITTLE scheduling) to 3.3\ms.

\subsubsection{Sustained Performance}

To characterize deployment viability, we run a 30-minute continuous 1080p playback validation on SM8650 (Adreno~750 + HTP~V75) with H.264 30\,fps content, 50\% brightness, WiFi on. Every frame pair is timed and logged (54,623 total). Full-frame logging adds CPU overhead ($\sim$30 writes/s, shell 45$^\circ$C vs.\ 41$^\circ$C under sampled logging, battery $-16$\% vs.\ $-12$\%); the figures below are therefore conservative upper bounds.

\textbf{Thermal phases.} The system exhibits three distinct thermal regimes: (1)~\emph{cold} (minutes~0--5): median 22.2\ms, HTP median 14.0\ms; (2)~\emph{warm steady state} (minutes~6--21): median 28.0\ms, HTP median 17.0\ms, as DVFS throttles the HTP from peak to sustained operating points; (3)~\emph{hot} (minutes~22--30): median 31.0\ms, HTP median 17.6\ms, with second-stage throttling under sustained thermal load.

\textbf{Frame drop statistics.} Over the full 30-minute run, 94.9\% of frame pairs complete within the 33.3\ms budget (note: this includes the favorable cold-start phase at minutes~0--5; the warm/hot steady-state rate is $\sim$94\%). The 5.1\% that exceed it are concentrated in the hot phase (minutes~22--30: 11\% drop rate vs.\ 2--3\% in warm steady state). Of the 2,795 over-budget frames, 80\% are isolated single-frame events attributable to OS scheduling jitter rather than sustained overload: the longest consecutive drop is 10 frames (0.33\,s), and the median burst length is 1. The 148 extreme outliers ($>$50\ms) are dominated by GPU scheduling spikes (GPU stage 20--40\ms vs.\ typical 3.7\ms), not HTP degradation.

\textbf{Budget analysis.} The 12.8\ms NPU-only latency (Table~\ref{tab:latency}) expands to 28.4\ms end-to-end once prealignment, copies, and synchronization are included; HTP increases from 12.8\ms (BURST) to 17.0\ms in-pipeline due to DVFS throttling. The warm-steady-state median of 28.0\ms leaves 5.3\ms of margin. Hot-phase degradation suggests a thermal-aware frame-skip policy could maintain artifact-free playback at the cost of reduced temporal enhancement ratio.

\section{Discussion}
\label{sec:discussion}

\textbf{Resolution-dependent residual gains.} The neural residual adds $+2.25\dB$ on Vimeo but only $+0.67\dB$ on Xiph 1080p (ANVIL-S vs.\ MV Blend). At higher resolution, prealignment already reduces residual motion substantially, leaving less room for neural refinement---consistent with the zero-parameter MV Blend itself improving from $+5.59\dB$ over naive blend on Vimeo to $+3.66\dB$ on Xiph.

\textbf{Perceptual and temporal quality.} Replacing differentiable warping with residual prediction produces smoother outputs, reflected in both LPIPS (0.148 vs.\ 0.077 on Xiph) and temporal fidelity. Table~\ref{tab:temporal} reports tOF (mean inter-frame optical flow magnitude via RAFT-Small~\cite{raft2020} at 540p) and warping error (WE) across 12 Xiph sequences.

\begin{table}[!htbp]
\centering
\caption{Temporal quality on Xiph 1080p (12 sequences, 15 consecutive pairs each). tOF: mean inter-frame optical flow magnitude (RAFT-Small at 540p). WE: warping error.}
\label{tab:temporal}
\renewcommand{\arraystretch}{1.15}
\begin{tabular}{lccc}
\toprule
Method & tOF & tOF$_\text{dev}$$\downarrow$ & WE$\downarrow$ \\
\midrule
GT              & 8.93 & ---  & 0.034 \\
RIFE HDv3       & 7.42 & \textbf{1.62} & \textbf{0.027} \\
ANVIL-S         & 6.91 & 2.09 & 0.027 \\
ANVIL-M         & 6.87 & 2.09 & 0.027 \\
MV Blend     & 6.13 & 2.83 & 0.026 \\
\bottomrule
\end{tabular}
\end{table}

RIFE achieves better temporal fidelity (tOF$_\text{dev}$ 1.62 vs.\ 2.09), but warping error is comparable across all methods (0.026--0.027), indicating that ANVIL is smoother than RIFE, not less correct. These gaps are the quantified cost of eliminating \texttt{grid\_sample} and iterative flow from the NPU graph.

\textbf{Encoding robustness.} Training uses per-triplet encoding (high-quality I-frame reference), while deployment extracts MVs from arbitrary streams. Sweeping preset, CRF, and B-frame configuration on Xiph 1080p shows \textbf{B-frames are the only significant factor} ($-1.16\dB$ for \texttt{bframes=3} vs.\ \texttt{bframes=0}); preset ($\pm 0.07\dB$) and CRF ($+0.12\dB$) are negligible because x264's motion estimation operates before quantization and the smoothing pipeline absorbs MV quality differences. For uncontrolled content with B-frames, the deployed system restricts interpolation to $d_\text{ref}=1$ frames, passing through frames where reliable MVs are unavailable.

\textbf{Codec scope.} ANVIL currently relies on H.264 MV side data via FFmpeg's \texttt{export\_mvs} (also supporting MPEG-1/2/4). HEVC and VP9/AV1 decoders lack MV export in FFmpeg---these are API gaps, not architectural constraints. Modern codecs (HEVC~\cite{hevcstandard2012}, VVC~\cite{vvcstandard2021}) provide richer motion models that could improve prealignment if per-block MVs become accessible through public mobile decoding APIs.

\textbf{Practical coverage.} Under controlled encoding (\texttt{bframes=0}), every inter-frame is interpolatable (30$\to$60\,fps). For uncontrolled content with \texttt{bframes=1} or 3, approximately 50\% of frames qualify (30$\to$45\,fps). The 16\% battery drain over 30 minutes (software decoding) is a further constraint relative to hardware decode.

\textbf{INT8 quantization implications.} Potential mitigations for iterative flow collapse---QAT~\cite{jacob2018quantization} or mixed-precision schemes keeping recurrent states in FP16---remain unexplored on current mobile NPU runtimes. More broadly, quantization-aware design should consider \emph{graph-level recurrent patterns}, not just individual operator sensitivity.

\section{Conclusion}
\label{sec:conclusion}

We presented ANVIL, a framework that addresses three structural barriers to deploying video frame interpolation on mobile NPUs---prohibitive \texttt{grid\_sample} latency, INT8 quantization collapse in iterative flow methods, and memory-bound operation dominance---by decomposing VFI into codec-side MV prealignment (CPU/GPU) and NPU-side residual refinement restricted to compute-bound operators. Per-operator INT8 analysis identifies Add-on-recurrent-state amplification as a key mechanism behind quantization collapse in iterative flow architectures, verified across two methods and two backends. ANVIL-M achieves 33.66\dB on Vimeo90K (0.6\dB below RIFE), the measured cost of NPU-deployable design; end-to-end system validation on SM8650 confirms 28.4\ms median VFI latency over 54,623 frame pairs during 30-minute playback (94.9\% within budget). The approach currently requires H.264 MV side data; extending to other codecs and uncontrolled content remains future work.

\section*{Acknowledgment}

The author gratefully acknowledges the individuals who provided mobile devices for cross-platform benchmarking. Claude (Anthropic) and ChatGPT (OpenAI) were used to assist with code development, including benchmark scripts, evaluation pipelines, and deployment code, as well as with English language editing. They were also used to assist in the interpretation of experimental and profiling results, including structural analysis of performance bottlenecks and profiling patterns. The author designed all experiments, directed all implementation, independently verified all analyses and results, wrote and approved the final manuscript, and takes sole responsibility for the scientific content of this paper.

\balance
\bibliographystyle{IEEEtran}
\bibliography{references}

\end{document}